\documentclass[11pt]{article}
\usepackage{amssymb}
\usepackage{epsfig}
\begin{document}
\title{ R-parity violation effect on the top-quark pair production at linear colliders
 \footnote{Supported
by National Natural Science Foundation of China.}} \vspace{3mm}
\author{{ Wang Lei$^{2}$, Ma Wen-Gan$^{1,2}$, Hou Hong-Sheng$^{2}$, Zhang Ren-You$^{2}$, and Sun Yan-Bin$^{2}$} \\
{\small $^{1}$ CCAST (World Laboratory), P.O.Box 8730, Beijing
100080, P.R.China} \\
{\small $^{2}$ Department of Modern Physics, University of Science and Technology} \\
{\small of China (USTC), Hefei, Anhui 230027, P.R.China}}
\date{}
\maketitle \vskip 20mm
\begin{abstract}
We investigate in detail the effects of the R-parity lepton number
violation in the minimal supersymmetric standard model (MSSM) on
the top-quark pair production via both $e^--e^+$ and
$\gamma-\gamma$ collision modes at the linear colliders. We find
that with the present experimental constrained $\rlap/{R}$
parameters, the effect from $\rlap/{R}$ interactions on the
processes $e^+e^-\to t\bar{t}$ and $e^+e^- \to \gamma\gamma \to
t\bar{t}$ could be significant and may reach $-30\%$ and several
percent, respectively. Our results show that the $\rlap/{R}$
effects are sensitive to the c.m.s. energy and the relevant
$\rlap/{R}$ parameters. However, they are not sensitive to squark
and slepton masses when $m_{\tilde{q}} \geq 400~GeV$ (or
$m_{\tilde{l}} \geq 300~GeV$) and are almost independent on the
$\tan\beta$.

\end{abstract}

\vfill \eject \baselineskip=0.36in
\renewcommand{\theequation}{\arabic{section}.\arabic{equation}}
\renewcommand{\thesection}{\Roman{section}}
\newcommand{\nb}{\nonumber}
\makeatletter      
\@addtoreset{equation}{section}
\makeatother       
\section{Introduction}
\par
The new physics beyond the standard model (SM) has been
intensively studied over the past years\cite{haber}. The minimal
supersymmetric standard model (MSSM) is currently the most popular
one among the extensions of the SM. The R-parity is defined as
\begin{equation}
R=(-1)^{2B+L+2S}
\end{equation}
where B, L and S represent the baryon number, lepton number and
intrinsic spin of the particle, respectively. This leads to a
discrete $Z_2$ symmetry in the MSSM lagrangian. For all the SM
particles they have $R=1$ and for all of their supersymmetric
partners $R=-1$. Usually we consider the R-parity is conserved in
the MSSM, but the most general superpotential consistent with the
gauge symmetry of the SM can introduce R-parity violating terms as
follow\cite{yin}
\begin{equation}
\label{eq1}
W_{\rlap/{R}} = \epsilon_{ij} (\lambda_{IJK}
\tilde{L}_i^I \tilde{L}_j^J
     \tilde{R}^K+\lambda_{IJK}' \tilde{L}_i^I \tilde{Q}_j^J \tilde{D}^K+
     \epsilon_I H_i^2 \tilde{L}_j^I)+\lambda_{IJK}''\tilde{U}^I
     \tilde{D}^J \tilde{D}^K,
\end{equation}
where $\tilde{L}^I, \tilde{Q}^I, H^I$ represent the SU(2) doublets
of lepton, quark and Higgs superfields, respectively, while
$\tilde{R}^I, \tilde{U}^I\, \tilde{D}^I$ are singlets of lepton
and quark superfields, I,J,K are flavor indices. All these terms
can lead to catastrophically high decay rate for proton. We must
require typically $\lambda'\lambda'' \leq 2 \times 10^{-26}$
\cite{sher1} in order to get a proton lifetime longer than
$10^{40}$s \cite{munich}. This is highly unnatural, unless either
one or both of $\lambda'$ and $\lambda''$ are identically zero. In
usual MSSM, we require R-parity conservation (i.e. $\lambda' =
\lambda'' =0$), it constrains the model more than what is really
necessary. $\lambda'=0$ or $\lambda''=0$ is quite adequate. In
most models motivated by unification (including supergravity),
there is a preference for allowing the lepton number violation
over the baryon number violation. In addition, the lowest
generation $\tilde{U}^I \tilde{D}^J \tilde{D}^K$ operators have
the strictest laboratory bounds, e.g. $\lambda_{121}'' \leq
10^{-6}$ \cite{sher2}. Because the quark mixing is not zero, it is
hard to construct models which allow for large coupling
$\lambda''$ and satisfy the strict constraint on
$\lambda_{121}''$. During the last few years experimental search
for the effects of $\rlap/{R}$ interactions has been done with
many efforts. Up to now we have only some upper limits on these
$\rlap/{R}$ parameters. It is necessary to continue these works on
finding $\rlap/{R}$ signal or getting further stringent
constraints on the $\rlap/{R}$ parameters in future experiments.

\par
Top quark was firstly discovered in 1995 by the CDF and D0 at the
Fermilab Tevatron. People believe that accurate measurement of top
quark pair production at the present and future colliders should
be possible in finding the physical effects beyond the SM. Any
deviation of observable in top quark pair production process from
the SM prediction would give a hint of new physics. Therefore,
testing the $\rlap/{R}$ effect on top pair production process is
an attractive task in high energy experiment.

\par
In previous studies, a lot of effort has been invested in the top
quark pair production at present and future colliders, such as the
CERN LEP2, CERN Large Hadron Collider (LHC), Tevatron, and the
proposed linear colliders (LC): NLC\cite{NLC}, JLC\cite{JLC},
TESLA\cite{TESLA} and CLIC\cite{CLIC}. A electron-positron LC can
be designed to operate in either $e^+e^-$ or $\gamma\gamma$
collision mode. $\gamma\gamma$ collision is achieved by using
Compton backscattered photons in the scattering of intense laser
photons on the initial polarized $e^+e^-$ beams\cite{Com}. At
these machines in both $e^+e^-$ and $\gamma\gamma$ collision
modes, a great number of top quark pairs can be produced
\cite{ma1}. The signature of this kind of event is much cleaner
than that produced at hadron colliders. Ref.\cite{hollik} presents
the calculation of the MSSM one-loop radiative corrections to
process $e^+e^-\to t\bar{t}$ and shows the relative difference
between the predictions of the MSSM and the SM is typically below
$10\%$. Ref.\cite{anindya} presents the effects of lepton number
violating interactions on top-quark pair production via $e^+e^-$
collisions. The NLO QCD corrections in the SM and MSSM to
$\gamma\gamma\to t\bar{t}$ have been discussed in detail in
Ref.\cite{ma2}, the corrections are about $10\%$ and of the order
$10^{-2}$, respectively. Ref.\cite{denner} demonstrates the SM
electroweak (EW) corrections to $\gamma\gamma \to t\bar{t}$ can
reach almost $10\%$ in the collision energy region close to the
threshold. Ref.\cite{li} gives the O($\alpha m_t^2/m_w^2$) Yukawa
corrections to the $e^+e^- \to \gamma\gamma\to t\bar{t}$ in the
SM, the general two-Higgs-doublet model (2HDM) as well as the
MSSM. The corrections are about a few percent in the SM but can be
bigger than $10\%$ in the MSSM. In Ref.\cite{zhou} the SUSY
EW-like corrections in the $R$-conserving MSSM to $\gamma\gamma\to
t\bar{t}$ are calculated, and the corrections are about a few
percent for $\gamma\gamma\to t\bar{t}$ and one percent for
$e^+e^-\to \gamma\gamma\to t\bar{t}$. In this paper, we study the
$\rlap/R$ lepton number violating effects on both the processes
$e^+e^-\to t\bar{t}$ and $e^+e^-\to \gamma\gamma \to t\bar{t}$ at
a LC. The paper is organized as follows. In Sec.II. we give the
relevant theory and Feynman diagrams. In Sec.III. we present the
analytical calculations. The numerical results and discussions are
described in Sec.IV. Finally, we give a short summary. In the
Appendix, the related $\rlap/R$ lepton number violating Feynman
rules are listed.

\par

\section{The Relevant Theory and Feynman Diagrams}
In this section we briefly review the theory of MSSM with
$\rlap/R$ lepton number violation. The most general form of the
superpotential in the MSSM can be written as\cite{roy}:
\begin{equation}
{\cal W} = {\cal W}_{MSSM}+{\cal W}_{\rlap/{R}},
\end{equation}
where ${\cal W}_{MSSM}$ represents the R-parity conserved term,
which can be written as
\begin{equation}
{\cal W}_{MSSM} = \mu \epsilon_{ij} H_i^1 H_j^2+\epsilon_{ij} l_I
H_i^1 \tilde{L}_j^I \tilde{R}^I-u_I (H_1^2 C^{JI \ast}
\tilde{Q}_2^J-
     H_2^2 \tilde{Q}_1^J)\tilde{U}^I-d_I (H_1^1 \tilde{Q}_2^I -
     H_2^1 C^{IJ} \tilde{Q}_1^J) \tilde{D}^I.
\end{equation}
The $\rlap/R$ superpotential part ${\cal W}_{\rlap/{R}}$ is
shown in Eq.(\ref{eq1}). The soft breaking terms can be expressed
as
\begin{equation}
\begin{array}{lll}
{\cal L}_{soft} &=& -m_{H^1}^2 H_i^{1\ast}H_i^1-m_{H^2}^2
H_i^{2\ast} H_i^2-
     m_{L^I}^2 \tilde{L}_i^{I\ast} \tilde{L}_i^I-m_{R^I}^2 \tilde{R}^{I\ast}
     \tilde{R}^I-m_{Q^I}^2 \tilde{Q}_i^{I\ast} \tilde{Q}_i^I \\
&& -m_{D^I}^2
     \tilde{D}^{I\ast} \tilde{D}^I-m_{U^I}^2 \tilde{U}^{I\ast} \tilde{U}^I
     + (m_1 \lambda_B \lambda_B+m_2 \lambda_A^i \lambda_A^i+m_3 \lambda_G^a
     \lambda_G^a+h.c.) \\
&& + \{B \mu \epsilon_{ij} H_i^1 H_j^2+B_I \epsilon_I
     \epsilon_{ij} H_i^2 \tilde{L}_j^I+\epsilon_{ij} l_{sI} H_i^1 \tilde{L}_j^I
     \tilde{R}^I \\
&& +d_{sI}(-H_1^1 \tilde{Q}_2^I+C^{IK} H_2^1 \tilde{Q}_1^K)
     \tilde{D}^I+u_{sI}(-C^{KI\ast} H_1^2 \tilde{Q}_2^I + H_2^2 \tilde{Q}_1^I)
     \tilde{U}^I \\
&& +\epsilon_{ij} \lambda_{IJK}^S \tilde{L}_i^I \tilde{L}_j^J
     \tilde{R}^K+\lambda_{IJK}^{S'}(\tilde{L}_i^I \tilde{Q}_2^J \delta^{JK}-
     \tilde{L}_2^I C^{JK} \tilde{Q}_1^J)\tilde{D}^K+\lambda_{IJK}^{S''}
     \tilde{U}^I \tilde{D}^J \tilde{D}^K \\
&& + h.c.\}.
\end{array}
\end{equation}
The bilinear term $\epsilon_{i j} \epsilon_I H_i^2 \tilde{L}_j^I$
is usually considered to be smaller than trilinear terms, so we
assume that they are negligible in our work . For the reason
mentioned in the introduction we only consider the lepton number
violation. This means that $\lambda'' = 0$. Then ${\cal
L}_{\rlap/{R}}$ can be written as
\begin{equation}
\begin{array}{lll}
{\cal L}_{\rlap/{R}} &=& -\lambda_{ijk}\left[ \bar{e}_k P_L
\nu_i \tilde{e}_{j L}
   + \bar{\nu}_i^{c} P_L e_j \tilde{e}_{k L}^{*} + \bar{e}_k P_L e_j \tilde{\nu}_{iL} \right]\\
   &-& \lambda^{'}_{ijk} \left[ \bar{d}_k P_L d_j \tilde{\nu}_{iL} + \bar{d}_k P_L \nu_i \tilde{d}_{j}
    + \bar{\nu}_{iL}^c P_L d_j \tilde{d}_{kR}^{*}
    - \bar{d}_k P_L e_i \tilde{u}_{jL} - \bar{d}_k P_L u_{j} \tilde{e}_{iL} - \bar{e}_i^{c} P_L u_j \tilde{d}_{kR}^{*} \right]\\
   &-& \epsilon_{\alpha\beta\gamma} \lambda_{ijk}^{''} \left[
   \bar{d}_j^{c\beta} P_R d_k^{\gamma}\tilde{u}_{iR}^{\alpha} + \bar{u}_i^{c \alpha} P_R d_j^{\beta} \tilde{d}_{kR}^{\gamma} + \bar{u}_i^{c \alpha} P_R d_k^{\gamma} \tilde{d}_{jR}^{\beta}
   \right] + h.c.
\end{array}
\end{equation}
There $\alpha,\beta,\gamma$ are color indices of quarks. From the
lagrangian we get the relevant Feynman rules which are listed in
Appendix. The Feynman diagrams of the process $e^+e^- \to
t\bar{t}$ are plotted in Fig.1. Fig.1(a-b) are the tree level
R-parity conserving and violating diagrams, respectively. Fig.2
shows the Feynman diagrams of subprocess $\gamma\gamma \to
t\bar{t}$. Fig.2(a) is tree level diagram. Fig.2(b-f) are vertex,
box and quartic coupling diagrams with $\rlap/R$ interactions.
Since the R-conserving SUSY EW-like one-loop correction diagrams
were already presented in Ref.\cite{zhou}, we shall not plot them
here. In Fig.2 the diagrams which can be obtained by exchanging
the initial photons are not shown.
\par

\section{Calculation}
In all our calculations we use the t'Hooft-Feynman gauge. In the
loop diagram calculation we adopt the definitions of one-loop
integral functions in Ref.\cite{s13}, and use the dimensional
reduction (DR) scheme\cite{copper} and on-mass-shell (OMS)
scheme\cite{ross} to do renormalization. The numerical calculation
of the vector and tensor loop integral functions can be traced
back to scalar loop integrals as shown in the Ref.\cite{s14}.
\par
\subsection{Calculation of the process $e^+e^-\to t\bar{t}$ }
\par
We denote the process of $t\bar{t}$ production via $e^+e^-$
collision as
\begin{equation}
e^-(p_1)+e^+(p_2) \to \overline{t}(k_1)+ t(k_2).
\end{equation}
where $p_1,p_2,k_1$ and $k_2$ are the momenta of the incoming and
outgoing particles, respectively. The differential Born cross
sections in the R-parity conserving MSSM, corresponding to the
diagrams Fig.1(a), can be written as
\begin{equation}
 d \sigma_{MSSM}=  dP \frac{N_c}{4} \sum_{spin}
 |A^{(a)}_{\gamma}(\hat s,\hat t,\hat u)+ A^{(a)}_Z(\hat s,\hat t,\hat u)|^2,
\end{equation}
where $N_c=3$, the summation is taken over the spin of the initial
and final states, and $dP$ denotes the two-particle phase space
element. The factor $1/4$ results from the average over the spins
of the incoming photons. The $A^{(a)}_{\gamma}$ and $A^{(a)}_Z$
represent the amplitudes of the photon and Z boson exchange
diagrams at tree level, respectively. The Mandelstam kinematical
variables are defined as
\begin{equation}
 \hat{s} = (p_1 + p_2)^2, \quad
  \hat t = (p_1-k_1)^2, \quad
  \hat u = (p_1-k_2)^2,
\label{eq:mandel}
\end{equation}

\par
In the MSSM with the $\rlap/R$ lepton number violation, the tree
level differential cross sections can be expressed as
\begin{equation}
 d {\hat \sigma_{\rlap/{R}}}(\hat s,\hat t, \hat u)=  dP \frac{N_c}{4}
 \sum_{spin} |A^{(a)}_{\gamma}(\hat s,\hat t, \hat u)+ A^{(a)}_Z(\hat s,\hat t, \hat u) +
 \sum\limits_{k=1}^{2,3}A^{(b)}_{\tilde{D}_k}(\hat s,\hat t, \hat u)|^2,\\
\end{equation}
where $A^{(b)}_{\tilde{D}_k},~(k=1,2,3)$ are the amplitudes
corresponding to diagrams in Fig.1(b). We give the explicit
expressions of $A^{(a)}_{\gamma}$, $A^{(a)}_{Z}$ and
$A^{(b)}_{\tilde{D}_k}$ as below

\begin{equation}
\begin{array}{lll}
A^{(a)}_{\gamma}(\hat s,\hat t,\hat u)  & = &Q_t e^2 [\bar{u}
(k_1)\gamma_{\mu} v(k_2)]\frac{-i} {\hat s}[\bar{v}
(p_2)\gamma_{\mu} u(p_1)]\\
A^{(a)}_{Z}(\hat s,\hat t,\hat u)  & = & \frac{e^2}{s_w^2
c_w^2}[\bar{u} (k_1)\gamma_{\mu}(\frac{P_L}{2} - \frac{2s_w^2}{3})
v(k_2)]\frac{i}
{\hat{s}-m_Z^2}[\bar{v}(p_2)\gamma_{\mu}(-\frac{P_L}{2} + s_w^2) u(p_1)]\\
A^{(b)}_{\tilde{D}_k}(\hat s,\hat t,\hat u)  & =
&-(\lambda^{\prime}_{13k})^2 \sum\limits_{j=1}^{2}\left\{ [\bar{u}
(k_2)Z_{D_k}^{2j}P_L u(p_1)] \frac{i}{\hat{u}-m_{\tilde{D}_{kj}}^2}
 [\bar{v} (p_2)Z_{D_k}^{2j}P_R v(k_1)]\right\}, \\
\end{array}
\end{equation}
where $Z_{D_k}^{ij}$ represents the elements of the matrix used to
diagonalize the down-type squark mass matrix, $k$ is the
generation index.
\par
\subsection{Calculation of the process $e^+e^- \to \gamma\gamma \to t\bar{t}$ }
In this subsection we present the calculation of the process
$e^+e^-\to\gamma\gamma\to t\bar{t}$. We denote the subprocess as
\begin{equation}
\gamma(p_1)+\gamma(p_2) \rightarrow \overline{t}(k_1)+ t(k_2),
\end{equation}
The Lorentz invariant matrix element at tree level for the process
$\gamma\gamma\to t\bar{t}$ can be written as
\begin{equation}
A_0=A^{(t)}_0(\hat s,\hat t,\hat u) + A^{(u)}_0(\hat s,\hat t,\hat u)
\end{equation}
where
\begin{equation}
\begin{array}{lll}
A^{(t)}_0(\hat s,\hat t, \hat u)  & = &- \frac{iQ_t^2 e^2}{\hat t
- m_t^2} \bar{u} (k_1)\gamma_{\mu}(m_t + \rlap/k_1 - \rlap/p_1)
\gamma_{\nu} v(k_2)\varepsilon^{\mu}(p_1)\varepsilon^{\nu}(p_2)
\end{array}
\end{equation}
\begin{equation}
\begin{array}{lll}
A^{(u)}_0(\hat s,\hat t,\hat u)  & = &- \frac{iQ_t^2e^2}{\hat u -
m_t^2} \bar{u} (k_1)\gamma_{\nu}(m_t + \rlap/k_1 -
\rlap/p_2)\gamma_{\mu}
v(k_2)\varepsilon^{\nu}(p_2)\varepsilon^{\mu}(p_1)
\end{array}
\end{equation}
The corresponding differential cross section is written as
\begin{equation}
\frac{d \hat{\sigma}_{0}(\hat{s},\hat{t},\hat{u})} {d \hat{t}}=
   \frac{1}{4}\frac{N_c}{16\pi\hat{s}^2}\sum_{spin} |A_0|^2,
\end{equation}
where the summation is taken over the spin of initial and final
states. The SUSY EW-like corrections in the R-parity conserving
MSSM to top pair production via photon-photon collisions were
calculated in Ref.\cite{zhou}. In this work we present only
one-loop contributions involving $\rlap/R$ interactions.
\par
The counterterm of top quark waves function $\delta Z_{t}$ is
decided by the one-particle irreducible two-point function
$i\Gamma(p^2)$ with the on-mass-shell (OMS) condition. The
renormalized $\rlap/R$ part of the top quark two-point function
can be defined as
\begin{equation}
\begin{array} {lll}
\hat{\Gamma}_{tt}(p^2) &=&
         (\rlap/p-m_{t}) +  \left [ \rlap/p P_{L}
    \hat{\Sigma}_{tt}^{L}(p^2)
   + \rlap/p P_{R} \hat{\Sigma}_{tt}^{R}(p^2)
   + P_{L} \hat{\Sigma}_{tt}^{S,L}(p^2)
   + P_{R} \hat{\Sigma}_{tt}^{S,R}(p^2) \right].
\end{array}
\end{equation}
The corresponding $\rlap/R$ parts of the unrenormalized
self-energies are
\begin{eqnarray}
\Sigma^{S,L}_{tt}(p^2) = 0,~~~  \Sigma^{S,R}_{tt}(p^2) = 0,~~~
 \Sigma^{R}_{tt}(p^2) = 0
\end{eqnarray}
\begin{equation}
\begin{array}{lll}
\Sigma^{L}_{tt}(p^2) = \frac{1}{16 \pi^2}
\sum\limits_{i=1}^{2,3}\sum\limits_{j=1}^{2}\sum\limits_{k=1}^{2,3}(|V_{\tilde{D}_{ij}E_kt}|^2B_0[p^2,
m_{\tilde{D}_{ij}}^2, m_{E_k}^2]\\
  + |V_{\tilde{D}_{ij}E_kt}|^2B_1[p^2, m_{\tilde{D}_{ij}}^2, m_{E_k}^2]
  + |V_{\tilde{E}_{kj}D_it}|^2B_1[p^2, m_{D_{i}}^2, m_{\tilde{E}_{kj}}^2])
\end{array}
\end{equation}
where $i$, $k$ are the generation indexes and $j$ is the sparticle
index. Using the OMS renormalization conditions\cite{ross}, we get
the renormalization constants as
\begin{equation}
\delta \Sigma_{tt}(p^2) = C_{L} \rlap/p P_{L} +
   C_{R} \rlap/p P_{R} - C^{L}_{S} P_{L} - C^{R}_{S} P_{R},
\end{equation}
The renormalization constants for the $t-t-\gamma$ vertex
$V_{tt\gamma}$ are
\begin{equation}
 \delta V_{\gamma tt} =
   -i e \gamma^{\mu}[ C^{L} P_{L} + C^{R} P_{R} ],
\end{equation}
where
\begin{eqnarray}
C_L &=& \frac{1}{2} (\delta Z^L_{tt} +
   \delta Z^{L\dag}_{tt}), \nonumber \\
C_R &=& \frac{1}{2} (\delta Z^R_{tt} +
   \delta Z^{R\dag}_{tt}), \nonumber \\
C^{L}_{S} &=& \frac{m_t}{2} (\delta Z^L_{tt} +
   \delta Z^{R\dag}_{tt}) +
   \delta m_t, \nonumber \\
C^{R}_{S} &=& \frac{m_t}{2}
   (\delta Z^R_{tt} +
   \delta Z^{L\dag}_{tt}) +
   \delta m_t.
\end{eqnarray}
\begin{eqnarray}
\delta m_t &=& \frac{1}{2} \tilde{Re}
  \left [ m_t \Sigma^{L}_{tt} (m_t^2) + m_t
  \Sigma^{R}_{tt}(m_t^2) +
  \Sigma^{S,L}_{tt}(m_t^2) +
  \Sigma^{S,R}_{tt}(m_t^2)
  \right ],
\end{eqnarray}
\begin{eqnarray}
\delta Z^{L}_{tt} &=& - \tilde{Re}\Sigma^{L}_{tt} (m_t^2) -
\frac{1}{m_t} \tilde{Re}
  \left [ \Sigma^{S,R}_{tt} (m_t^2)
  - \Sigma^{S,L}_{tt} (m_t^2)
  \right ] \nonumber \\
&-& m_t \frac{\partial}{\partial p^2} \tilde{Re}
  \left \{ m_t \Sigma^{L}_{tt}(p^2)
  + m_t \Sigma^{R}_{tt}(p^2)
  \right. \nonumber \\
&+& \left. \left.  \Sigma^{S,L}_{tt}(p^2) +
     \Sigma^{S,R}_{tt}(p^2) \right\} \right|_
     {p^2=m_t^2},
\end{eqnarray}
\begin{eqnarray}
\delta Z^{R}_{tt} &=& -\tilde{Re}\Sigma^{R}_{tt} (m_t^2) -
  m_t \frac{\partial} {\partial p^2} \tilde{Re}
  \left\{ m_t \Sigma^{L}_{tt}(p^2) +
  m_t \Sigma^{R}_{tt}(p^2) \right. \nonumber \\
&+& \left. \left. \Sigma^{S,L}_{tt}(p^2) +
    \Sigma^{S,R}_{tt}(p^2) \right \} \right|_
    {p^2=m_t^2},
\end{eqnarray}
where $\tilde{Re}$ takes the real part of the loop integrals. We
use ${A}^{v}$, ${A}^{b}$, ${A}^{q}$, ${A}^{s}$ and $A^{ct}$ to
represent the amplitude parts contributed by vertex, box, quartic
, self-energy diagrams (shown in Fig.2) and counterterms,
respectively. The renormalized matrix elements from the one-loop
diagrams are written as
\begin{equation} \begin{array}{lll}
\delta {A}_{1-loop} &=&
    {A}^{v}+{A}^{b}+{A}^{q}+{A}^{self}+A^{ct}\\
&=& \epsilon_{\mu}(p_1)\epsilon_{\nu}(p_2) \bar{u}(k_1) \left\{
f_1g_{\mu\nu} + f_2g_{\mu\nu}\gamma_5 + f_3k_{2\nu}\gamma_\mu
+ f_4g_{\mu\nu}\rlap/{p_2} + f_5k_{2\nu}\gamma_5\gamma_\mu \right. \\
&+& \left. f_6 g_{\mu\nu}\gamma_5\rlap/{p_2} +
f_7\gamma_\mu\gamma_\nu + f_8k_{2\nu}\gamma_\mu\rlap/{p_2} +
f_9\gamma5\gamma_\mu\gamma_\nu +
f_{10}k_{2\nu}\gamma_5\gamma_\mu\rlap/{p_2}
+ f_{11}\gamma_\mu\gamma_\nu\rlap/{p_2}\right. \\
&+& \left. f_{12}\gamma_5\gamma_\mu\gamma_\nu\rlap/{p_2} +
f_{13}k_{2\mu}k_{2\nu}+ f_{14}k_{2\mu}k_{2\nu}\gamma_5 +
f_{15}k_{2\mu}\gamma_\nu+ f_{16}k_{2\mu}\gamma_5\gamma_\nu\right. \\
&+& \left. f_{17}k_{2\mu}\gamma_\nu\rlap/{p_2} +
f_{18}k_{2\mu}\gamma_5\gamma_\nu\rlap/{p_2} +
f_{19}k_{2\mu}k_{2\nu}\rlap/{p_2} +
f_{20}k_{2\mu}k_{2\nu}\gamma_5\rlap/{p_2} \right\} v(k_2),
\end{array} \end{equation}
where $f_i~(i=1 \sim 20)$ are the form factors. Then we get the
one-loop corrections to the cross section from the $\rlap/R$ part:
\begin{equation}
 \Delta \hat{\sigma}_{1-loop}(\hat{s}) = \frac{N_c}{16 \pi \hat{s}^2}
             \int_{\hat{t}^{-}}^{\hat{t}^{+}} d\hat{t}~
        2 Re{\overline{\sum\limits_{spin}^{}}}\left( {A}_{0}^{\dag} \cdot
        \delta {A}_{1-loop} \right),
\end{equation}
where $\hat{t}^\pm = (m_t^2-\frac{\hat{s}}{2}) \pm
\frac{\hat{s}}{2} \sqrt{1-4 m_t^2/\hat{s}}$, and the bar over the
summation means average over initial spins. With the integrated
photon luminosity in the $e^{+}e^{-}$ collision, the total cross
section of the process $e^+e^- \to \gamma\gamma\to t\bar{t}$ can
be written as
\begin{eqnarray}
\label{integration}
\hat\sigma(s)= \int_{E_{0}/ \sqrt{s}} ^{x_{max}} d z \frac{d%
{\cal L}_{\gamma\gamma}}{d z} \hat{\sigma}(\gamma\gamma \to
t\overline{t} \hskip 3mm
 at \hskip 3mm  \hat{s}=z^{2} s)
\end{eqnarray}
with $E_{0}=2m_t$, and $\sqrt{s}$($\sqrt{\hat{s}}$) being the
$e^{+}e^{-}$($\gamma\gamma$) center-of-mass energy.  $\frac{d\cal
L_{\gamma\gamma}}{d z}$ is the distribution function of photon
luminosity, which is defined as:
\begin{eqnarray}
\frac{d{\cal L}_{\gamma\gamma}}{dz}=2z\int_{z^2/x_{max}}^{x_{max}}
 \frac{dx}{x} F_{\gamma/e}(x)F_{\gamma/e}(z^2/x)
\end{eqnarray}
For the initial unpolarized electrons and laser photon beams, the
energy spectrum of the back scattered photon is given by
\cite{photon spectrum}
\begin{eqnarray}
\label{structure}
F_{\gamma/e}=\frac{1}{D(\xi)}[1-x+\frac{1}{1-x}-
\frac{4x}{\xi(1-x)}+\frac{4x^{2}}{\xi^{2}(1-x)^2}]
\end{eqnarray}
where
\begin{eqnarray}
D(\xi)=\left(1-\frac{4}{\xi}-\frac{8}{{\xi}^2}\right)\ln{(1+\xi)}+\frac{1}{2}+
  \frac{8}{\xi}-\frac{1}{2{(1+\xi)}^2}, \\ \nonumber
  \xi=\frac{4E_0 \omega_0}{{m_e}^2},
\end{eqnarray}
$m_{e}$ and $E_{0}$ are the incident electron mass and energy,
respectively, $\omega_0$ is the laser-photon energy, and $x$ is the
fraction of the energy of the incident electron carried by the
backscattered photon. In our calculation, we choose $\omega_0$
such that it maximizes the backscattered photon energy without
spoiling the luminosity via $e^{+}e^{-}$ pair creation. Then we
have ${\xi}=2(1+\sqrt{2})$, $x_{max}\simeq 0.83$, and
$D(\xi)=1.8$.

\par

\section{Numerical results and discussion}
We take the input parameters as $m_e=0.511~MeV$,
$m_\mu=105.66~MeV$, $m_\tau=1.777~GeV$, $m_Z = 91.188~GeV$, $m_W =
80.41~GeV$, $m_u = 5~MeV$, $m_c = 1.35~GeV$, $m_t = 174.3~GeV$,
$m_d = 9~MeV$, $m_s = 150~MeV$, $m_b = 4.3~ GeV$, $\alpha_{EW} =
1/128$\cite{pdg}. Because our numerical results show that the
$\rlap/R$ corrections are almost independent on $\tan\beta$, we
take $\tan\beta = 4$ as a representative selection in the
$\rlap/R$ case. The scalar fermion mass terms in lagrangian are
written as
\begin{equation}
\label{eq3}
 -{\cal L}_{M_{\tilde{f}}} = \left(
\tilde{f}_L^{*},\tilde{f}_R^{*}\right) {\cal M}_{\tilde{f}}^2
\left(
\begin{array}{c} \tilde{f}_L \\ \tilde{f}_R \end{array} \right) = \left(
\tilde{f}_L^{*},\tilde{f}_R^{*}\right) \left( \begin{array}{cc}
{\cal M}_{\tilde{f}~LL}^2 & {\cal M}_{\tilde{f}~LR}^2 \\
{\cal M}_{\tilde{f}~LR}^{2*} & {\cal M}_{\tilde{f}~RR}^2
\end{array} \right)
\left( \begin{array}{c} \tilde{f}_L \\ \tilde{f}_R \end{array}
\right),
\end{equation}
The matrix of the scalar fermion mass is
\begin{eqnarray}
\label{mixsf} && {\cal M}_{\tilde{f}}^2 = \left(
\begin{array}{cc}
M_{\tilde{F}_L}^2 + m_f^2 + \cos 2\beta(I_3^{fL}-Q_f s_W^2)M_Z^2 & m_f (A_f - \mu \kappa_f ) \\
m_f (A_f - \mu \kappa_f )^{*} & M_{\tilde{F}^{'}}^2 + m_f^2 + \cos
2\beta Q_f s_W^2 M_Z^2
\end{array} \right), \\
&& ( \kappa_f, m_{\tilde{F}_L}^2, m_{\tilde{F}^{'}}^2 ) =\left\{
\begin{array}{cl}
(\cot \beta, m_{\tilde{Q}}^2, m_{\tilde{U}}^2 ),  & \hbox{when $\tilde{f}$ is up-squark} \\
(\tan \beta, m_{\tilde{Q}}^2, m_{\tilde{D}}^2 ),  & \hbox{when $\tilde{f}$ is down-squark} \\
(\tan \beta, m_{\tilde{E}_L}^2, m_{\tilde{E}_R}^2 ),  & \hbox{when
$\tilde{f}$ is slepton}
\end{array} \right.
\end{eqnarray}
where $Q_f$ is the charge of scalar fermion, $I_3^{fL}$ is the
third component of the left-hand fermion weak isospin,
$m_{\tilde{Q}}, m_{\tilde{U}}, m_{\tilde{D}}, m_{\tilde{E}_L},
m_{\tilde{E}_R}$ are the soft-SUSY-breaking masses and $A_f$ is
the soft-SUSY-breaking trilinear coupling parameter. If we do not
consider the CP violation, the matrix elements are real and can be
diagonalized as
\begin{equation}
R^{\tilde{f}} {\cal M}_{\tilde{f}}^2 R^{\tilde{f} \dag} =
\hbox{diag}\{m_{\tilde{f}_1}^2 , m_{\tilde{f}_2}^2 \},
\end{equation}
The mass eigenstates of scalar fermions can be obtained from the
transformation of the current eigenstates,
\begin{eqnarray}
\left( \begin{array}{c} \tilde{f}_1 \\ \tilde{f}_2 \end{array}
\right) &=& R^{\tilde{f}}
  \left( \begin{array}{c} \tilde{f}_L \\ \tilde{f}_R \end{array} \right) =
  \left( \begin{array}{cc}
    \cos \theta_{\tilde{f}} & \sin \theta_{\tilde{f}} \\
   -\sin \theta_{\tilde{f}} & \cos \theta_{\tilde{f}}
  \end{array} \right)
  \left( \begin{array}{c} \tilde{f}_L \\ \tilde{f}_R \end{array} \right),  \\
\label{sangle} \tan 2 \theta_{\tilde{f}} &=& \frac{2 {\cal
M}_{\tilde{f}~LR}^2 }
       {{\cal M}_{\tilde{f}~LL}^2 - {\cal M}_{\tilde{f}~RR}^2},\\
 m_{\tilde{f}_{1,2}}^2 &=&  \frac{1}{2}  \left\{
        {\cal M}_{\tilde{f}~LL}^2 + {\cal M}_{\tilde{f}~RR}^2 \mp
       \sqrt{({\cal M}_{\tilde{f}~LL}^2 - {\cal M}_{\tilde{f}~RR}^2)^2 + 4 ({\cal M}_{\tilde{f}~LR}^2)^2 }
\right \}.
\end{eqnarray}
If we take $\theta_{\tilde{f}}$ as an input parameter, then we get
\begin{eqnarray}
\label{eq2}
 m_{\tilde{f}_{1,2}} &=&  \sqrt{\frac{1}{2}  \left\{
        {\cal M}_{\tilde{f}~LL}^2 + {\cal M}_{\tilde{f}~RR}^2 \mp
       |{\cal M}_{\tilde{f}~LL}^2 - {\cal M}_{\tilde{f}~RR}^2   |/\cos{2 \theta_{\tilde{f}}}
\right\}}.
\end{eqnarray}
\par
In our following numerical calculation, we set $\mu= 200~GeV$ and
$\tan\beta=4$. In the squark sector, we follow the way in choosing
input parameters presented in Ref.\cite{zhou} and assume
$m_{\tilde{Q}^{1,2}} = m_{\tilde{U}^{1,2}} $= $m_{\tilde{D}^{1,2}}
= M_Q$ for the first and second generations, $m_{\tilde{Q}^{3}} =
m_{\tilde{U}^{3}} $= $m_{\tilde{D}^{3}} = M_{Q^3}$ for the third
generation, and $\theta_{\tilde{t}}=44.325^{\circ}$,
$\theta_{\tilde{u}, \tilde{d}, \tilde{c}, \tilde{s},
\tilde{b}}=0$. In the slepton sector, we assume
$m_{\tilde{E}_L^{\alpha}} = m_{\tilde{E}_R^{\alpha}} =
A_{e^\alpha} = M_L$, where $e^1, e^2, e^3$ are e, $\mu$, $\tau$,
respectively.

\par
In the numerical calculation for the process $e^+e^-\to t\bar{t}$,
we take the input data of squark sector describing above and
$\{M_Q, M_{Q^3}\} = \{300, 200\}~GeV,~\{500, 300\}~GeV,~\{800,
500\}~GeV$, respectively. In the calculation of the process
$e^+e^-\to\gamma\gamma\to t\bar{t}$, besides the input parameters
of squark sector mentioned before, we take $M_{Q^3}=200~GeV$ and
$\{M_Q, M_L\} = \{300, 150\}~GeV$, $\{500, 150\}~GeV$, $\{300,
600\}~GeV$, $\{500, 600\}~GeV$ respectively  or otherwise stated.
Then the masses of physical squarks and sleptons are obtained from
Eqs.(\ref{eq3})-(\ref{eq2}). According to the experimental upper
limits of the coupling parameters in the $R$-parity violating
interactions presented in Ref.\cite{B.A}, we take the relevant
$\rlap/R$ parameters as $\lambda^{'}_{131} = 0.05$,
$\lambda^{'}_{133} = 0.002$, $\lambda^{'}_{233} = 0.2$,
$\lambda^{'}_{132} = \lambda^{'}_{231}$ = $\lambda^{'}_{232}$ =
$\lambda^{'}_{331}$ =$ \lambda^{'}_{332} = \lambda^{'}_{333} =
0.4$ for numerical representation.

\par
The cross section of the process $e^+e^- \to t\bar{t}$ as a
function of $\sqrt{\hat{s}}$ is plotted in Fig.3, where $R$-parity
conserving (RC) and R-parity violating (RV) results are presented.
The four curves correspond to (1) R-parity conserving case,
(2) R-parity violating case with $M_Q  = 300~GeV$ and $M_{Q^3} = 200~GeV$,
(3) R-parity violating case
with $M_Q = 500~GeV$ and $M_{Q^3} = 300~GeV$, (4) R-parity
violating case with $M_Q = 800~GeV$ and $M_{Q^3} = 500~GeV$,
respectively. We can see that the $\rlap/R$ effect on the
production cross section decreases with the increments of $M_Q$
and $M_{Q^3}$, and if the masses of squarks are about $200~GeV$
and $300~GeV$ and $\sqrt{\hat{s}} = 500~GeV$, the relative
$\rlap/R$ correction ($\delta=\frac{\Delta \sigma}{\sigma_0}$) can
reach about $-30\%$. So we can say that if $\rlap/R$ really
exists, the $\rlap/R$ effect on the cross section of the process
$e^+e^- \to t\bar{t}$ can be observed or its accurate measurement
can provide more stringent constraints on the masses of squarks,
sleptons and $\lambda^{'}$.

\par
Fig.4-8 demonstrate the $\rlap/R$ and $R$-conserving SUSY EW-like
one-loop corrections to $e^+e^- \to \gamma\gamma \to t\bar{t}$.
Fig.4(a) presents the dependence of the $\rlap/R$ and
$R$-conserving SUSY EW-like one-loop corrections $\Delta\sigma$ on
the $\sqrt{\hat{s}}$ for the subprocess $\gamma\gamma\to
t\bar{t}$. Fig.4(b) presents the dependence of the $\rlap/R$ and
$R$-conserving SUSY EW-like one-loop relative corrections
$\delta=\Delta\sigma/\sigma_{tree}$ on the $\sqrt{\hat{s}}$. In
Fig.4(a-b), we have $m_{\tilde{Q}^{3}} =200~GeV$. The solid line
is for $M_Q = 300~GeV, M_L=150~GeV$, the dashed line is for $M_Q =
500~GeV, M_L=150~GeV$. The dotted and dash-dotted lines are for
$M_Q = 300~GeV, M_L=600~GeV$ and $M_Q = 500~GeV, M_L=600~GeV$,
respectively, and dash-dot-dotted line for the corresponding
$R$-conserving SUSY EW-like corrections. On all the curves in both
Fig.4(a) and Fig.4(b), we can see line structures with small
spikes due to the resonance effects. For example, on some of the
curves in Fig.4(a) there exist small resonance spikes in the
region around the vicinities of $\sqrt{\hat{s}} \sim 2
m_{\tilde{b}_2}\sim 415~GeV$ and $\sqrt{\hat{s}}\sim
2m_{\tilde{b}_1}\sim 403~GeV$. On the curves of $M_Q = 300~GeV,
M_L=150~GeV$ and $M_Q = 300~GeV, M_L=600~GeV$, we can see other
resonance effect at $\sqrt{\hat{s}}\sim
2m_{\tilde{d}_1,\tilde{s}_1}\sim 602~GeV$ and $\sqrt{\hat{s}}\sim
2m_{\tilde{d}_2,\tilde{s}_2}\sim 610~GeV$. For the curves of $M_Q
= 300~GeV$, $M_L=600~GeV$ and $M_Q = 500~GeV$, $M_L=600~GeV$, the
resonance effect can be seen around the position of
$\sqrt{\hat{s}}\sim
2m_{\tilde{e}_{1,2},\tilde{\mu}_{1,2},\tilde{\tau}_{1,2}}\sim
1203~GeV$. The corresponding line structures due to resonance
effect are shown again in Fig.4(b). The $\rlap/{R}$ effects shown
in these two figures are very obvious, all the curves
corresponding to the different input $M_L$ and $M_Q$ values
demonstrate that the $\rlap/{R}$ corrections are comparable to the
$R$-conserving SUSY one-loop EW-like corrections. Especially for
the curves with $\{M_Q = 300~GeV$, $M_L = 150~GeV\}$ and $\{M_Q =
500~GeV$, $M_L = 150~GeV\}$, the $\rlap/{R}$ corrections are
larger than the corresponding one-loop $R$-conserving SUSY one and
the $\rlap/{R}$ relative correction for $\{M_Q = 300~GeV,~M_L =
150~GeV\}$ can reach $-4.1\%$.

\par
In Fig.5 and Fig.6 we take $M_{Q^3} = 200~GeV$,
$\sqrt{\hat{s}}=~500 GeV$ and depict the $\rlap/R$ relative
corrections as the functions of $M_Q$ and $M_L$ respectively. In
Fig.5 the curves correspond to $M_L=150~GeV$, $300~GeV$, $500~GeV$
and $800 ~GeV$, respectively, and $M_Q$ varies from $300~GeV$ to
$900~GeV$. We can see from Fig.5 that the corrections are
sensitive to the value of $M_Q$ when squark mass parameter $M_Q$
is below $400~GeV$, typically when $M_Q=300~GeV$ the relative
correction can reach $-3.0\%$. While when $M_Q$ is larger than
$400~GeV$, the relative corrections are not sensitive to $M_Q$ due
to the decouple theorem. In Fig.6 we choose $M_Q= 300~GeV$,
$500~GeV$ and $800~GeV$ respectively, $M_L$ goes from $100~GeV$ to
$900~GeV$. The spikes at $M_L \sim 245~GeV$ are from the resonance
effect because the relation of $\sqrt{\hat{s}}\sim
2m_{\tilde{e}_{1,2},\tilde{\mu}_{1,2},\tilde{\tau}_{1,2}}\sim
500~GeV$ exists. If $M_L$ is below $200~GeV$, for all taken $M_Q$
values in the figure, the corrections are relative large and can
reach $-3.0\%$. But when $M_L>300~GeV$, the $\rlap/R$ corrections
are not sensitive to $M_L$.

\par
In Fig.7 we have $\sqrt{\hat{s}} = 500~GeV$, $M_{Q^3} = 200~GeV$
and take (1)$M_Q = 300~GeV, M_L=150~GeV$ (solid line);  (2)$M_Q =
500~GeV, M_L=150~GeV$(dashed line); (3)$M_Q = 300~GeV,
M_L=600~GeV$(dotted line) and (4)$M_Q = 500~GeV,
M_L=600~GeV$(dash-dotted line), respectively, assuming
$\lambda^{'}_{132}=\lambda^{'}_{231}=\lambda^{'}_{232}=\lambda^{'}_{331}
=\lambda^{'}_{332}=\lambda^{'}_{333}=\lambda'$. The relative
$\rlap/R$ corrections to the subprocess $\gamma\gamma \to
t\bar{t}$ as the functions of $\lambda'$ are plotted in this
figure. We can see that the $\rlap/{R}$ corrections are negative
and reduce the cross section of the subprocess. The figure shows
that the relative corrections are getting larger with the
increment of the $\lambda'$ value when sleptons have small masses,
but the $\rlap/{R}$ effects would be very weak if sleptons are
heavy. When we have $\lambda' = 0.6$ and $M_Q = 300~GeV,
M_L=150~GeV$, the relative correction can reach $-6.6\%$.

\par
Fig.8(a) shows the one-loop $\rlap/R$ corrections to the parent
process $e^+e^-\to\gamma\gamma\to t\bar{t}$ as the functions of
the electron-positron colliding energy. We take again the input
parameter sets as: (1)$M_Q = 300~GeV, M_L=150~GeV$ (full line);
(2)$M_Q = 500~GeV, M_L=150~GeV$(dashed line); (3)$M_Q = 300~GeV,
M_L=600~GeV$(dotted line) and (4)$M_Q = 500~GeV,
M_L=600~GeV$(dash-dotted line), respectively, and vary the
$e^+e^-$ colliding energy from 0 to 2 TeV. At the position of
$\sqrt{s} \sim 1.2~TeV$ the absolute corrections reach their
maximal values, e.g., for the curve of $M_Q = 300~GeV,
M_L=150~GeV$, the curve has the maximal correction $\Delta
\sigma_{max}=-15.5~fb$. When $\sqrt{s}>1.2~TeV$, the absolute
corrections decrease with the increment of the colliding c.m.s
energy. Fig.8(b) shows the one-loop $\rlap/R$ and $R$-conserving
SUSY electroweak relative corrections with $M_{Q^3} = 200~GeV$ as
the functions of the colliding $e^+e^-$ energy. We can see that
when the colliding energy is below $1.2~TeV$, the $\rlap/R$
absolute relative corrections decrease apparently with the
increment of $\sqrt{s}$ except the curve with heavy masses of
sleptons and the first and second generation squarks ($M_Q =
500~GeV, M_L=600~GeV$). But when $\sqrt{s}$ is larger than
$1.2~TeV$, the $\rlap/R$ relative corrections are not very
sensitive to the c.m.s energy $\sqrt{s}$. In comparison with the
$R$-conserving one-loop SUSY electroweak corrections, we can
conclude that the $\rlap/{R}$ relative corrections are comparable
or even larger than the $R$-conserving one-loop SUSY electroweak
relative corrections for almost all the input parameters we used
in these two figures. When $M_Q = 300~GeV, M_L=150~GeV$, the
relative correction can reach $-3.6\%$.

\section{ Summary}
\par
In this paper, we studied the effect of the $R$-parity lepton
number violation in the MSSM on both important processes
$e^+e^-\to t\bar{t}$ and $e^+e^- \to \gamma\gamma \to t\bar{t}$ at
a LC. To the former process, we find that the $\rlap/R$ effect is
obviously related to the masses of squark and slepton. The heavier
the squarks and sleptons are, the smaller the $\rlap/R$ effect is.
The $\rlap/R$ relative correction can reach $-30\%$ with the
favorable parameters. To the second process, our calculation shows
that the $\rlap/R$ corrections to either subprocess
$\gamma\gamma\to t\bar{t}$ or parent process
$e^+e^-\to\gamma\gamma\to t\bar{t}$ are strongly related to the
colliding energy. The $\rlap/R$ relative correction can reach
several percent to both cross sections of the subprocess and
parent process. Although the $\rlap/R$ correction is smaller than
QCD correction\cite{ma2}, it can be even larger than the
$R$-conserving SUSY electroweak correction with suitable
parameters\cite{denner} \cite{li} \cite{zhou}. So the $\rlap/R$
effect on both processes could be significant and could be
measured experimentally, if the $\rlap/R$ really exists. We also
investigate the dependence of the $\rlap/R$ correction on the
relevant $\rlap/R$ input parameters, such as $M_L$, $M_{Q}$,
$M_{Q^3}$, $\lambda^{'}_{ijk}$, etc. We find that the $\rlap/R$
correction is strongly related to the input parameters $M_{Q}$,
$M_{Q^3}$, $M_L$ and $\lambda^{'}_{ijk}$ in some parameter space,
but is not sensitive to squark mass (or slepton mass) when
$m_{\tilde{q}} \geq 400~GeV$ (or $m_{\tilde{l}} \geq 300~GeV$) and
is almost independent on $\tan\beta$.

\par
\noindent{\large\bf Acknowledgments:} This work was supported in
part by the National Natural Science Foundation of China and a
grant from the University of Science and Technology of China.

\section{Appendix}
The relevant Feynman rules of R-parity violating interactions are
showed as below
\par
\begin{tabular}{ll}
\hspace{20mm} \epsfig{file=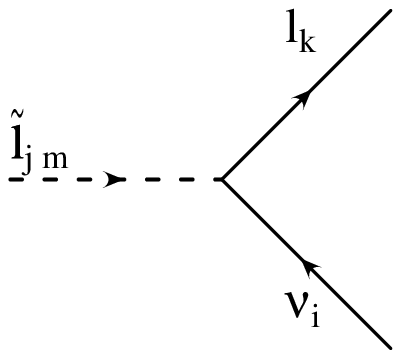, height=1.2 in, width=1.8 in}
\end{tabular} $-i \lambda_{ijk} R^{\tilde{e}_j}_{m1}P_L$  \\
\par
\begin{tabular}{ll}
\hspace{20mm} \epsfig{file=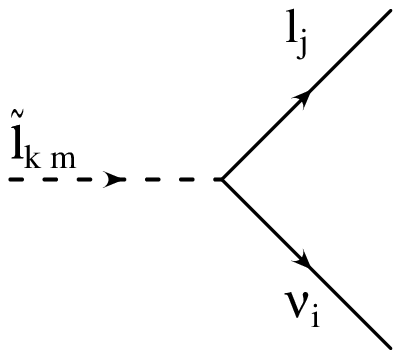, height=1.2 in, width=1.8 in}
\end{tabular} $-i \lambda_{ikj}R^{\tilde{e}_k}_{m2}P_R~C $  \\
\par
\begin{tabular}{ll}
\hspace{20mm} \epsfig{file=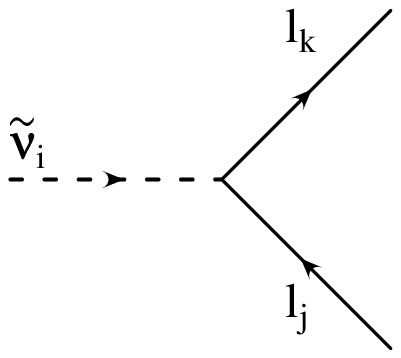, height=1.2 in, width=1.8 in}
\end{tabular} $-i \lambda_{ijk} P_L $  \\
\par
\begin{tabular}{ll}
\hspace{20mm} \epsfig{file=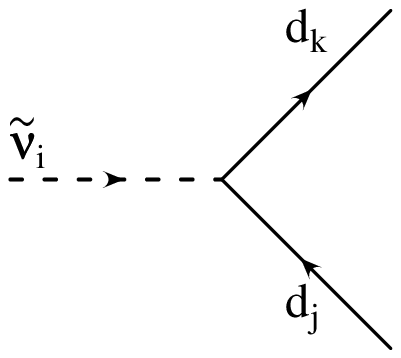, height=1.2 in, width=1.8 in}
\end{tabular} $-i \lambda_{ijk}^{'} P_L $  \\
\par
\begin{tabular}{ll}
\hspace{20mm} \epsfig{file=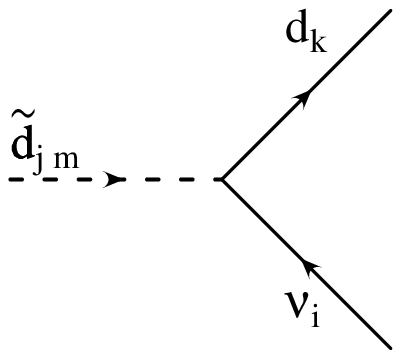, height=1.2 in, width=1.8 in}
\end{tabular} $-i \lambda_{ijk}^{'} R^{\tilde{d}_j}_{m1} P_L $  \\
\par
\begin{tabular}{ll}
\hspace{20mm} \epsfig{file=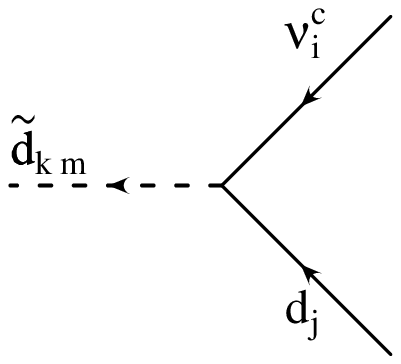, height=1.2 in, width=1.8 in}
\end{tabular} $i \lambda_{ijk}^{'} R^{\tilde{d}_k *}_{m2}C^{-1}~P_L $  \\

\par
\begin{tabular}{ll}
\hspace{20mm} \epsfig{file=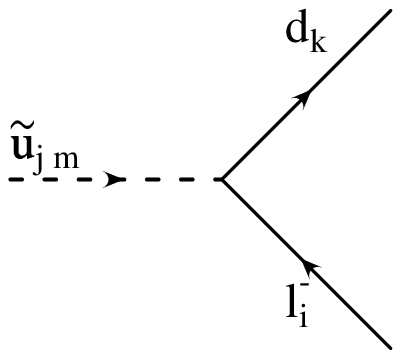, height=1.2 in, width=1.8 in}
\end{tabular} $i \lambda_{ijk}^{'} R^{\tilde{u}_j}_{m1} P_L $  \\
\par
\begin{tabular}{ll}
\hspace{20mm} \epsfig{file=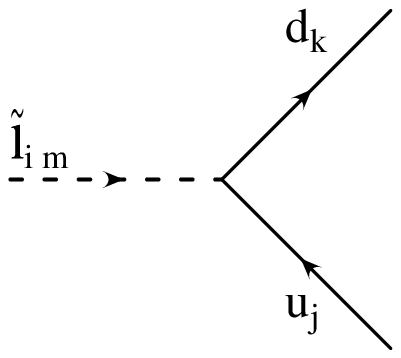, height=1.2 in, width=1.8 in}
\end{tabular} $i \lambda_{ijk}^{'} R^{\tilde{e}_i}_{m1} P_L $  \\
\par
\begin{tabular}{ll}
\hspace{20mm} \epsfig{file=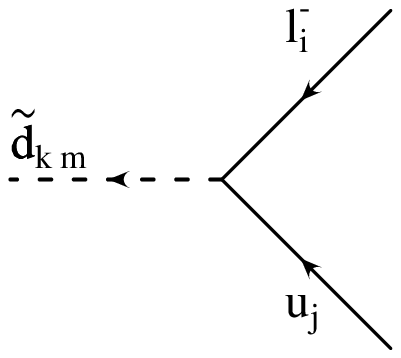, height=1.1 in, width=1.8 in}
\end{tabular} $-i \lambda_{ijk}^{'} R^{\tilde{d}_k *}_{m2} C^{-1} P_L $  \\
\par
\begin{tabular}{ll}
\hspace{20mm} \epsfig{file=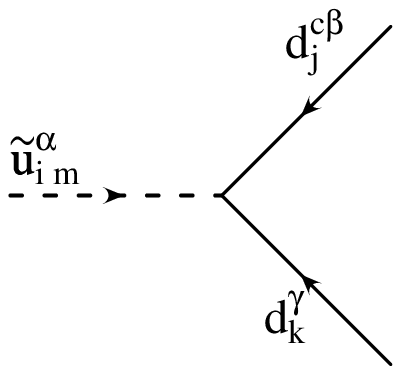, height=1.1 in, width=1.8 in}
\end{tabular} $i \varepsilon_{\alpha\beta\gamma} \lambda_{ijk}^{''} R^{\tilde{u}_i}_{m2} C^{-1}~P_R $  \\
\par
\begin{tabular}{ll}
\hspace{20mm} \epsfig{file=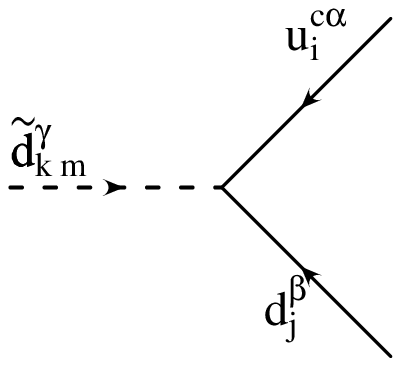, height=1.1 in, width=1.8 in}
\end{tabular} $i \varepsilon_{\alpha\beta\gamma} \lambda_{ijk}^{''} R^{\tilde{d}_k}_{m2} C^{-1}~P_R $  \\

\vskip 10mm
\begin{flushleft} {\bf Figure Captions} \end{flushleft}
\par
{\bf Fig.1}  The relevant Feynman diagrams for the process
$e^+e^-\to t\bar{t}$ in the MSSM at the tree-level: (a)the Feynman
diagrams for R-parity conserved MSSM part; (b)the Feynman diagrams
for R-parity violation MSSM part.
\par
{\bf Fig.2} The relevant Feynman diagrams for the subprocess
$\gamma\gamma\to t\bar{t}$ in the MSSM with $R$-parity lepton
number violation at the tree-level and the one-loop level diagrams
with $\rlap/R$ interactions: Fig.2(a) is tree level diagram.
Fig.2(b.1)-(b.8) are vertex diagrams. Fig.2(c.1)-(c.6) are box
diagrams. Fig.2(d.1)-(d.2) and Fig.2(f.1)-(f.2) are self-energy
diagrams. Fig.2(e.1)-(e.2) are quartic coupling diagrams.
\par
{\bf Fig.3} The cross section of the process $e^+e^- \to t\bar{t}$
as a function of $\sqrt{\hat{s}}$.
\par
{\bf Fig.4(a)} The one-loop $\rlap/{R}$ and $R$-conserving SUSY
EW-like corrections $\Delta\sigma$ as the functions of
$\sqrt{\hat{s}}$ for the subprocess $\gamma\gamma\to t\bar{t}$.
\par
{\bf Fig.4(b)} The one-loop $\rlap/{R}$ and $R$-conserving SUSY
EW-like relative corrections $\delta=\Delta\sigma/\sigma_{tree}$
as the functions of $\sqrt{\hat{s}}$ for the subprocess
$\gamma\gamma\to t\bar{t}$.
\par
{\bf Fig.5} The one-loop $\rlap/{R}$ relative corrections to the
subprocess $\gamma\gamma\to t\bar{t}$ as the functions of $M_Q$
with $\sqrt{\hat{s}} = 500~GeV$.
\par
{\bf Fig.6} The one-loop $\rlap/{R}$ relative corrections to the
subprocess $\gamma\gamma\to t\bar{t}$ as the functions of $M_L$
with $\sqrt{\hat{s}} = 500~GeV$.
\par
{\bf Fig.7} The one-loop $\rlap/{R}$ relative corrections to the
subprocess $\gamma\gamma\to t\bar{t}$ as the functions of
$\lambda'$ with  $\sqrt{\hat{s}} = 500~GeV$ (We assume
$\lambda^{'}_{132}=\lambda^{'}_{231}=\lambda^{'}_{232}=\lambda^{'}_{331}=\lambda^{'}_{332}=\lambda^{'}_{333}=\lambda'$).
\par
{\bf Fig.8(a)} The one-loop $\rlap/{R}$ corrections $\Delta\sigma$
to the parent process $e^+e^-\to\gamma\gamma\to t\bar{t}$ as the
functions of the c.m.s energy of the incoming electron-positron
pair.
\par
{\bf Fig.8(b)} The one-loop $\rlap/{R}$ and $R$-conserving SUSY
EW-like relative corrections to the parent process $e^+e^-\to\gamma\gamma\to t\bar{t}$ as the
functions of the c.m.s energy of the incoming electron-positron
pair.

\end{document}